\documentclass[10pt, conference, compsocconf]{IEEEtran}
\usepackage{cite}
\usepackage{color}
\usepackage{xspace}
\usepackage{amsmath}
\usepackage{array}
\usepackage{mdwmath}
\usepackage{mdwtab}
\usepackage{eqparbox}
\usepackage{paralist}
\usepackage{amsfonts}
\usepackage{bm}
\usepackage{algorithm}% http://ctan.org/pkg/algorithms                         
\usepackage{algpseudocode}                                                     
\usepackage[algo2e,linesnumbered,ruled,vlined]{algorithm2e}
\usepackage{caption}
\usepackage{subcaption}
\usepackage{fixltx2e}
\usepackage{float}
\usepackage{url}
\usepackage{tikz,pgfplots,pgfplotstable}
\usepackage{adjustbox}
\usepackage{booktabs}
\usetikzlibrary{pgfplots.groupplots}
\usetikzlibrary{arrows, backgrounds, calc, hobby, positioning}
\usetikzlibrary{shapes.geometric}
\pgfplotsset{compat=1.9}
\usepackage{mdframed}
\newmdtheoremenv{obs}{Observation}
\newcommand{\kpgm}[0]{{\em KPGM}\xspace}
\newcommand{\KPGM}[0]{{\em KPGM}\xspace}

\newcommand{\ERGM}[0]{{\em ERGM}\xspace}
\newcommand{\bter}[0]{{\em BTER}\xspace}
\newcommand{\mkpgm}[0]{{\em mKPGM}\xspace}
\newcommand{\mKPGM}[0]{{\em mKPGM}\xspace}
\newcommand{\tkpgm}[0]{{\em tKPGM}\xspace}
\newcommand{\tKPGM}[0]{{\em tKPGM}\xspace}
\newcommand{\xkpgm}[0]{{\em xKPGM}\xspace}
\newcommand{\xKPGM}[0]{{\em xKPGM}\xspace}
\begin{document}
\title{Modeling Graphs Using a Mixture of Kronecker Models}
\author{
  \IEEEauthorblockN{Suchismit Mahapatra}
\IEEEauthorblockA{Computer Science and Engineering\\
State University of New York at Buffalo\\
suchismi@buffalo.edu}
%\and
%\IEEEauthorblockN{Amey Sanjay Mahajan}
%\IEEEauthorblockA{Computer Science and Engineering\\
%State University of New York at Buffalo\\
%Buffalo, NY\\
%ameysanj@buffalo.edu}
\and
  \IEEEauthorblockN{Varun Chandola}
\IEEEauthorblockA{Computer Science and Engineering\\
State University of New York at Buffalo\\
chandola@buffalo.edu}
}

\maketitle
%\pagenumbering{arabic}
%\setcounter{page}{1}%Leave this line commented out.

\begin{abstract}
Generative models for graphs are increasingly becoming a popular tool for researchers to generate realistic approximations of graphs. While in the past, focus was on generating graphs which follow general laws, such as the power law for degree distribution, current models have the ability to learn from observed graphs and generate synthetic approximations. The primary emphasis of existing models has been to closely match different properties of a single observed graph. Such models, though stochastic, tend to generate samples which do not have significant variance in terms of the various graph properties. We argue that in many cases real graphs are sampled drawn from a graph population (e.g., networks sampled at various time points, social networks for individual schools, healthcare networks for different geographic regions, etc.). Such populations typically exhibit significant variance. However, existing models are not designed to model this variance, which could lead to issues such as overfitting. We propose a graph generative model that focuses on matching the properties of real graphs and the natural variance expected for the corresponding population. The proposed model adopts a mixture-model strategy to expand the expressiveness of Kronecker product based graph models (KPGM), while building upon the two strengths of KPGM, viz., ability to model several key properties of graphs and to scale to massive graph sizes using its elegant fractal growth based formulation. The proposed model, called {\bf x-Kronecker Product Graph Model}, or \xkpgm, allows scalable learning from observed graphs and generates samples that match the mean and variance of several salient graph properties. We experimentally demonstrate the capability of the proposed model to capture the inherent variability in real world graphs on a variety of publicly available graph data sets.

%\boldmath
\end{abstract}
\section{Introduction}
\label{sec:introduction}
Data occurs as graphs and networks in a wide variety of applications, ranging from social sciences to biology. Graph analysis methods are required to understand the structural properties of graphs. One important class of graph analysis methods deal with finding {\em generative} mechanisms and models that generate graphs with such structural properties. A crucial application of these models is to generate synthetic graphs which ``match'' the structural properties of real world graphs. Given the limited availability of real world graph data, mainly because of high cost and privacy concerns, such synthetic (and anonymized) graphs are a valuable resource for researchers to understand network behavior in domains such as systems analysis (e.g., the Internet), bioinformatics, and social sciences, while allowing for anonymity and privacy~\cite{Casas-Roma:2013}. Applications include understanding malware propagation in social networks~\cite{Sanzgiri:2013}, understanding fraud in healthcare~\cite{Chandola:2013}, etc. Graph generative models also allow researchers to produce realistics simulations at desired scale which are vital to understand issues such as handling scalability challenges and modeling temporal evolution.

What should be the salient characteristics of such a generative model? {\em First}, it should be able to capture the properties of real-world graphs. Traditionally the focus has been on general ``laws''  that are expected to be obeyed by real world graphs, such as power laws for degree distributions, small  diameters, communities, etc. However, to accurately represent the real world graph, matching on local graph properties such as edges, transitive triangles, etc., is important. {\em Second}, the model should be able to scale to massive graph sizes, to be applicable in domains where graphs tend to be big. {\em Third}, to allow generation (simulation) of any sized synthetic graphs, the model should be parametric and should allow learning of the parameters from one or more observed graphs.

\begin{figure}[h]
  \centering
  \includegraphics[width=\linewidth]{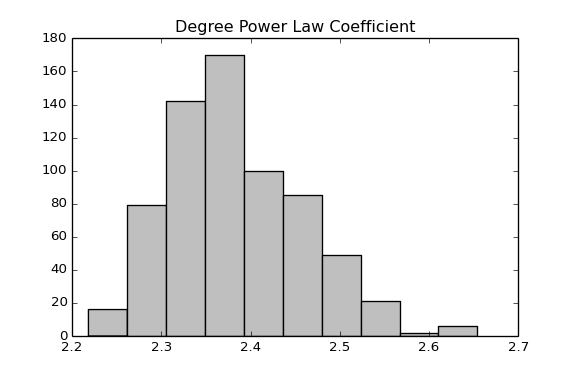}
  \caption{Illustration of the variance in power law coefficient ($\alpha$) for a population of over 700 Autonomous Systems (AS) graphs sampled at different time points. Few anomalous samples were removed from the original population.}
  \label{fig:variance}
\end{figure}

Table~\ref{tab:comparisontable} compares several existing graph generative models in terms of which of the above characteristics. Note that several models satisfy all three characteristics. However, we argue that there is an additional characteristic that needs to be incorporated into the generation models -- the ability to {\bf model the natural variance} in a population of graphs~\cite{Moreno:2010}. Real world graphs can be thought of as being generated from a natural process. Graph models {\em emulate} this process to generate similar synthetic graphs. We argue that graphs that are generated by the same process exhibit a natural variance in terms of the structural properties, and hence synthetic graphs generated from a model should mirror the same variance. Examples of such populations include, graphs collected at different times, social networks for different groups of people (e.g., schools), healthcare networks for different spatial regions, etc. Moreno et al., empirically demonstrated such variance in several social and email network graph populations~\cite{Moreno:2010}. Figure~\ref{fig:variance} illustrates the variance in the degree power law coefficient ($\alpha$) for a population of Autonomous Systems (AS) graphs. Observe that $\alpha$ exhibits bounded variance and the empirical distribution is close to the normal distribution. Similar behavior is observed for many other graph properties (See Figure~\ref{fig:sampledistributions1}).

%Most modern world applications deal with massive sized graphs. For most applications, Graphs and networks have long been one of the most natural form in which data has existed in a wide variety of applications, ranging from social sciences to biology. Graph analysis methods are required to understand the structure, interaction, and dynamics in such graphs. As sensing and storage capabilities evolve, the sizes of graphs in most domains are becoming increasingly larger. This has motivated the need for methods which can scale to such large graph sizes.
\begin{table*}[h]
\centering
{\footnotesize
\begin{tabular}{|l|c|c|c|c|c|c||c|}
\hline
& PA~\protect\cite{Barabasi:1999} & ERGM~\protect\cite{Wasserman:1996} & CL~\protect\cite{Aiello:2001}& \bter~\protect\cite{Seshadhri:2011a}& \kpgm~\protect\cite{Leskovec:2010}& \mkpgm~\protect\cite{Moreno:2013}& {\bf \xkpgm}\\
\hline
1. {\em Learnable}              & $\times$& $\surd$ & See note& See note& $\surd$ & $\surd$ & $\surd$\\
2. {\em Scalable Learning}      & --      & $\surd$ &   --    & --      & $\surd$ & $\times$& $\surd$\\
3. {\em Scalable Generation}    & $\times$& $\times$& $\surd$ & $\surd$ & $\surd$ & $\surd$ & $\surd$\\
4. {\em Match Local Properties} & $\times$& $\times$& $\times$& $\surd$ & $\surd$ & $\surd$ & $\surd$\\
5. {\em Capture Variance}       & $\times$& $\times$& $\times$& $\times$& $\times$& $\surd$ & $\surd$\\
%5. {\em Multiple Modes} & $\times$& $\surd$& & & $\times$& $\times$& $\times$& $\surd$\\
\hline
\end{tabular}
}
\caption{Comparison of existing graph generation models with the proposed \xkpgm model. {\em PA - Preferential Attachment, ERGM - Exponential Random Graph Models, CL - Chung-Lu, \bter - Block Two Erd\"{o}s-R\'{e}nyi, \kpgm - Kronecker Product Graph Model, \mkpgm - Mixed KPGM, \xkpgm - Proposed Model}. {\em Note:} The CL and \bter models can be used to generate a synthetic graph from a given real graph with approximately the same number of nodes, however it does not allow generation of arbitrary sized synthetic graphs.}
\label{tab:comparisontable}
\end{table*}

In summary, the research community needs a scalable graph generative model which matches properties of real graphs, including the variance exhibited by a population of graphs. However, as shown in Table~\ref{tab:comparisontable}, most existing graph generators lack in at least one of the desired characteristics listed above.

In particular, the {\bf Kronecker Product Graph Model} (\kpgm)~\cite{Leskovec:2010} has been shown to satisfy some of the above properties. The multiplicative nature of the model allows for fast sampling of massive sized graphs and has been shown, both analytically and empirically, to generate graphs that follow many power-law characteristics for several global graph properties. Moreover, a {\em method of moments} based approach was proposed for parameter learning with \kpgm~\cite{Gleich:2012}, which allows it to scale to massive graphs. However, recent papers have identified few shortcomings of \kpgm.

The single biggest argument made {\em against} \KPGM is that they lack the ability to capture the natural variability observed in real world graphs~\cite{Moreno:2010, Seshadri:2013}. Researchers have shown that the synthetic graphs sampled from \KPGM show little variation in terms of several graph properties.
%This issue is significant because it means that a \KPGM model learnt from a single graph cannot {\em represent} (i.e., assign a high probability score) to a different graph observed in the same domain. 
The problem is attributed to the fractal nature of growth in the generative model. The {\bf tied-KPGM} (\tKPGM) and the {\bf mixed-KPGM} (\mKPGM) are two variants proposed to alleviate the issues with \KPGM~\cite{Moreno:2010}. However, as we demonstrate experimentally, these next generation models are also not {\em expressive} enough to model the natural variance in the data. Moreover, the estimation process using the {\em simulated method of moments} approach is expensive and not scalable to learning from massive graphs.

To alleviate the lack of variation in existing models, we propose a mixture based approach for graph generation using Kronecker product. %One possible explanation for the lack of variance exhibited by \kpgm is the fact that at every time step, the growth of the graph is the same as previous step. To alleviate that issue, we introduce the idea of using multiple initiator matrices  and randomly use one of them at every iteration to grow the existing graph. On one hand, the proposed model, referred to as {\bf x-Kronecker Product Graph Model} (or \xkpgm), has the same desirable scalable behavior (for both learning and generation) as \kpgm. On the other hand, by introducing the randomness in the generative process, the proposed model not only matches the real graph properties but also models the population variance more effectively than \mkpgm.
%This approach allows significantly better modeling of graphs expected to be derived from same distribution, i.e., random subgraphs. Results on several real world graphs show that the proposed model outperforms the existing Kronecker product based models.
%\subsection{Our Contributions}
In particular, this paper makes the following specific contributions:
\begin{enumerate}
\item We propose a mixture-model based Kronecker product graph model (\xkpgm). 
%We refer to the proposed model as \xKPGM to maintain nomenclature continuity with existing methods (\KPGM, \tKPGM, and \mKPGM). 
We also show that the existing models are particular instantiations of the proposed \xKPGM model.
\item We analytically examine the expected properties of the graph generated by the proposed model.
\item We derive expressions for the expectactions for several salient graph properties for our model. Using these expressions, we provide a {\em method of moments} based parameter learning algorithm for the proposed model. 
\item We propose a random subgraph based method to evaluate generative models for graphs.
\item We show, analytically and empirically, that \xKPGM matches the real graph properties and captures the natural variability in graphs more effectively than existing Kronecker models and other graph generative models.
\end{enumerate}

\section{Related Work}
\label{sec:related}
Generative models for graphs has been a widely studied field for many decades. One of the earliest ones is the {\em Erd\H{o}s-R\'{e}nyi} model~\cite{Renyi:1960} which has obvious shortcomings as it fails to capture properties of real-world graphs. More recent models have focused on the {\em preferential attachment} property of nodes, i.e., new nodes tend to form edges with nodes with greater degree~\cite{Barabasi:1999}. However, in such models, the graph is grown one node at a time, which makes them inherently serial and unscalable. Lately, there has been an emphasis on models which can be learnt from observed graphs. These include the {\bf Exponential Random Graph Models} (\ERGM) (also referred to as p* models)~\cite{Wasserman:1996}. \ERGM essentially defines a log linear model over all possible graphs $G$, $p(G|\theta) \propto exp(\theta^Ts(G)$, where $G$ is a graph, and $s$ is a set of functions, that can be viewed as summary statistics for the structural features of the network. Another popular and well-known network models are {\em Stochastic Block Models} (SBM)~\cite{Snijders:1997} in which each node belongs to a cluster and the relationships between nodes are determined by their cluster membership. While conventional SBM are defined for non-overlapping community assignments, many overlapping or mixed variants have also been introduced~\cite{Airoldi:2008}. Another relevant model is the {\em Chung-Lu (CL) model}~\cite{Aiello:2001} in which the probability of an edge is proportional to the product of the degrees of its end vertices. While CL model effectively captures the degree distribution, it performs poorly for other properties such as clustering coefficient. A recent extension to CL model, called the {\em Block Two-Level Erd\H{o}s-R\'{e}nyi} (\bter) model~\cite{Seshadhri:2011a} is shown to match both the degree distribution and clustering coefficient on several graphs. However the \bter model is not truly generative as it only allows for creation of a synthetic graph which is exactly the same size as the observed graph. For generating arbitrary sized graphs one needs to provide parameters instead of learning them.

Kronecker product based generative models are increasingly becoming popular~\cite{Leskovec:2010}. However, given their limited expressiveness~\cite{Moreno:2010, Seshadri:2013}, several variants have been proposed~\cite{Moreno:2010,Moreno:2013a}. The most promising extension, \mkpgm, is able to capture the variance in a graph population, however, the parameter estimation phase is expensive. We will be discussing the original model and two variants in more detail in the paper. There also have been other related papers that improve \kpgm in other ways, e.g., the Multiplicative Attribute Graph Models~\cite{Kim:2012}.

\section{Background}
\label{sec:background}
In this section we introduce the various Kronecker product based models that have been proposed for modeling large graphs. For clarity, we will use the notation used in this section for the subsequent sections of the paper.
%\subsection{Graph Properties}
%\label{subsec:properties}
%One can measure several different properties of a graph, both at global and local levels. There have been several properties that have typically been considered by existing graph generative models. We use some of these key properties for parameter estimation of our proposed model as well as to evaluate the performance. One of the key property is the {\em degree distribution}. Most real world graphs follow the power law with respect to the degree distribution, i.e., $n_k \propto k^{-\alpha}$, where $\alpha > 0$ is the power law exponent and $n_k$ is the number of nodes with degree $k$. The second property is the {\em diameter} or the 
\subsection{Kronecker Product Graph Model}
\label{subsec:kpgm}
\kpgm~\cite{Leskovec:2010} is a fractal growth model. A graph $\mathcal{G}$ with $N$ nodes is obtained by first generating a $N\times N$ stochastic matrix (containing entries between 0 and 1). The edge between node $i$ and $j$ is independently (from other edges) generated using a Bernoulli trial with the $ij^{th}$ entry from the matrix. The generative algorithm starts with an initial matrix $\mathcal{P}_1 = \Theta$, which is $b\times b$; typically $b$ is set to 2 or 3, e.g.:
\begin{equation}
\mathcal{P}_1 = \Theta = 
\left[
  \begin{array}{cc}
    \theta_{11} & \theta_{12}\\
    \theta_{21} & \theta_{22}
  \end{array}
\right]
\label{eqn:0}
\end{equation}
The algorithm takes repeated Kronecker product of $\mathcal{P}_1$ with itself to generate a larger matrix. For example,
\begin{equation}
\mathcal{P}_n = \underbrace{\mathcal{P}_1 \otimes \mathcal{P}_1 \otimes \ldots \otimes \mathcal{P}_1}_{n \text{ times}}
\label{eqn:1}
\end{equation}
Note that the matrix $\mathcal{P}_n$ will have $b^{n}$ rows (and columns). This matrix is then used to generate a graph $\mathcal{G} = ({\bf V},{\bf E})$ with nodes ${\bf V} = \{1,2,\ldots,N\}$. For each pair $(u,v)$, where $1 \le u,v \le b^n$, a Bernoulli sample is generated with parameter $\mathcal{P}_n[u,v]$. If the sample is 1 (success), edge $(u,v)$ is added to ${\bf E}$. We denote this process as a {\em realization} of $\mathcal{P}_n$ to get the adjacency matrix, $A$ ($=R(\mathcal{P}_n)$).

The \kpgm model can assign a probability to a given graph $\mathcal{G}$ as long as the correspondence of nodes in $\mathcal{G}$ to the rows of $\mathcal{P}_n$ are given (denoted by $\sigma$). Given the \kpgm model ($\mathcal{P}_1 = \Theta$) and the correspondence function $\sigma$, the probability of an observed graph $\mathcal{G} = ({\bf V},{\bf E})$ is given by:
\begin{equation}
P(\mathcal{G}|\Theta,\sigma) = \prod_{(u,v)\in {\bf E}}\mathcal{P}_n[\sigma(u),\sigma(v)]\prod_{(u,v) \notin {\bf E}}(1-\mathcal{P}_n[\sigma(u),\sigma(v)])
\label{eqn:2}
\end{equation}
where $\mathcal{P}_n$ is derived from $\Theta$ using~\eqref{eqn:1}.

Given an observed graph $\mathcal{G}$, two approaches for learning the model parameters ($\Theta$) have been proposed. The first is an MLE approach that approximates computes the gradient of the log-likelihood of the graph (See~\eqref{eqn:2})~\cite{Leskovec:2010}. Since the correspondence function $\sigma$ is unknown, the learning algorithm searches over the factorial possible permutations using {\em Metropolis-Hastings sampling} and then uses a gradient descent approach to update the parameters in $\Theta$. The second learning approach is based on {\em method of moments}~\cite{Gleich:2012}. This method avoids the issue of searching over the factorial space of permutations.

It has been shown that \kpgm generates graphs which match real networks in terms of several properties such as skewed degree distribution, short network path length, etc. Moreover, the model allows very fast sampling of large graphs ($O(\vert{\bf E}\vert)$). However, recent work has identified several limitations with \kpgm. Seshadri et al.~\cite{Seshadri:2013} have shown that graphs generated from KPGM have 50-75\% isolated vertices. Moreno et al.~\cite{Moreno:2010} have observed, both analytically and empirically, that graphs generated from \kpgm do not capture the variability observed in the real networks expected to be generated from the same source. They have attributed this shortcoming to:
\begin{inparaenum}[(i)]
\item use of independent edge probabilities,
\item fractal expansion, and
\item small number of model parameters.
\end{inparaenum}
In fact, after several Kronecker products, most entries in the eventual stochastic matrix tend to become homogeneous. This results in a lack of variance since all sampled graphs appear similar.

Moreno et al.~\cite{Moreno:2010} investigated several simple approaches to induce variability in \kpgm generated graphs. These include using larger initiator matrices and sampling $\Theta$ from a distribution instead of using a point estimate. None of these variations induced significant variability in the sampled graphs. The same authors make the following key observation:
\begin{obs}
For a graph $\mathcal{G} = ({\bf V},{\bf E})$ generated by \kpgm, $Var(\vert{\bf E}\vert) \le \mathbb{E}[\vert{\bf E}\vert]$, independent of $n$.
\label{obs:1}
\end{obs}
The authors empirically observed that in real networks, the estimated variance of the number of edges across multiple variations of the same graph (across time) is significantly greater than the mean. Thus \kpgm cannot generate graphs with large variance.
\subsection{Tied Kronecker Product Graph Model}
\label{subsec:tkpgm}
Moreno et al.~\cite{Moreno:2010} proposed a generalization of \kpgm that allows larger variance in the properties of the graphs sampled from the model by inducing edge dependence in the generation process. In the generative process, an adjacency matrix is {\em realized} after each iteration from the current matrix $\mathcal{P}_t$, denoted as $R(\mathcal{P}_t)$. The next stochastic matrix, $\mathcal{P}_{t+1}$ is obtained by performing a Kronecker product of the realized adjacency matrix and the initiator matrix, i.e., $\mathcal{P}_{t+1} = R(\mathcal{P}_t) \otimes \mathcal{P}_1$. Note that for \kpgm, only one realization is done (after the $n^{th}$ iteration). Thus, the final adjacency matrix can be obtained using recursive realizations and Kronecker products, i.e.,
\begin{equation}
A = \underbrace{R(\ldots R(R(\mathcal{P}_1) \otimes \mathcal{P}_1) \ldots)}_{n \text{ realizations}}
\label{eqn:5}
\end{equation}
The above model is called tied \kpgm or \tkpgm since the realization at every step can also be thought of as {\em tying} the Bernoulli parameters. The authors show analytically that the variance for the expected number of edges in graphs generated using \tkpgm is higher than \kpgm, in fact the variance was higher than desired which motivated the variant discussed next.
\subsection{Mixed Kronecker Product Graph Model}
\label{subsec:mkpgm}
The mixed \kpgm or \mkpgm variation allows the graph to grow as \kpgm (without realizations) for first $l$ steps and then acts as \tkpgm for remaining $n-l$ steps. \mkpgm has an additional parameter $l$ which denotes the extent of tying in the model. Note that for $l = 1$, \mkpgm is equivalent to \tkpgm and for $l = n$, \mkpgm is equivalent to \kpgm.

\section{Natural Variability in Real Graphs}
\label{sec:variability}
A vital requirement for any graph generative model is the ability of the model to capture the variability across multiple observed samples. In fact, both \tkpgm and \mkpgm were motivated due to the limited ability of previous models to capture this variability. Most generative graph models have focused on capturing the properties observed for a single instance. But such models are not equipped to model a set of graphs, assumed to be sampled from the same statistical distribution. In many real world applications, one can obtain multiple samples of graphs which would exhibit variance in the different properties. For example, a set of social networks of students for different colleges, or a set of citation networks collected for different decades or disciplines. We show one example in Figure~\ref{fig:variance} where we consider daily AS networks across several years. In general, obtaining such populations to study the variance is challenging.

In this paper, we address this challenge by generating random subgraphs of a given graph. For a given large graph, the idea is to extract subgraphs using a sampling strategy such that the properties of the original graph are preserved in the subgraph. Several such methods have been proposed in the literature~\cite{Leskovec:2006,Hubler:2008,Sethu:2012}. In this paper we use a variant of the {\em forest fire} model to generate subgraphs.  In this method, a random node is chosen and the neighborhood of the chosen node is traversed in a breadth first approach. Each outgoing edge from the node is added to the sample with a certain probability. The nodes at the end of the chosen edges are then further ``burnt'' and the ``fire'' is further spread.%The second approach is to consider graphs representing the same system but observed over time. 

To illustrate the variability we consider several publicly available real world graphs listed in Table~\ref{tab:datasets}. %The last row refers to a collection of 733 graphs representing daily instances of the Autonomous systems (AS) from November 8, 1997 to January 2, 2000. We used this collection to understand the variability across time.
\begin{table}
\centering
{\footnotesize
\begin{tabular}{|c|p{1.1in}|p{0.28in}|p{0.32in}|}
\hline
Name & Description & Nodes & Edges\\
\hline
as~\cite{Snap} &CAIDA AS Relationship Graph & 6,474& 13,233\\
ca-astroPh~\cite{Snap} & Collaboration network of Arxiv Astro Physics & 18,772 & 396,160\\
elegans~\cite{Duch:2005} & C.~elegans metabolic network & 453 & 4,596\\
hep-ph~\cite{Snap} & Citation network from Arxiv HEP-PH & 34,546 & 421,578\\
netscience~\cite{Newman:2006} & Coauthorship network of scientists & 1,589 & 5,484\\
protein~\cite{Jeong:2001} & Protein interaction network for Yeast & 1,870 & 4,480\\
\hline
\end{tabular}
}
\caption{Publicly available graph data sets used in the evaluation. More details available at \protect\url{http://www.cse.buffalo.edu/~chandola/research/bigdata2015graphs/info.html}.}
\label{tab:datasets}
\end{table}
For each graph in Table~\ref{tab:datasets}, we sample 200 subgraphs (number of nodes were typically 4 times less than the original number of nodes) using the {\em forest-fire} approach and measure characteristics of each subgraph. The empirical distribution for several characteristics are shown in Figure~\ref{fig:sampledistributions1}. We study the following five important characteristics:
\begin{inparaenum}[(i)]
\item power law coefficient ($\alpha$) of the degree distribution
\item number of edges,
\item number of triangles,
\item average path length, and
\item average clustering coefficient.
%\item greatest component size.
\end{inparaenum}
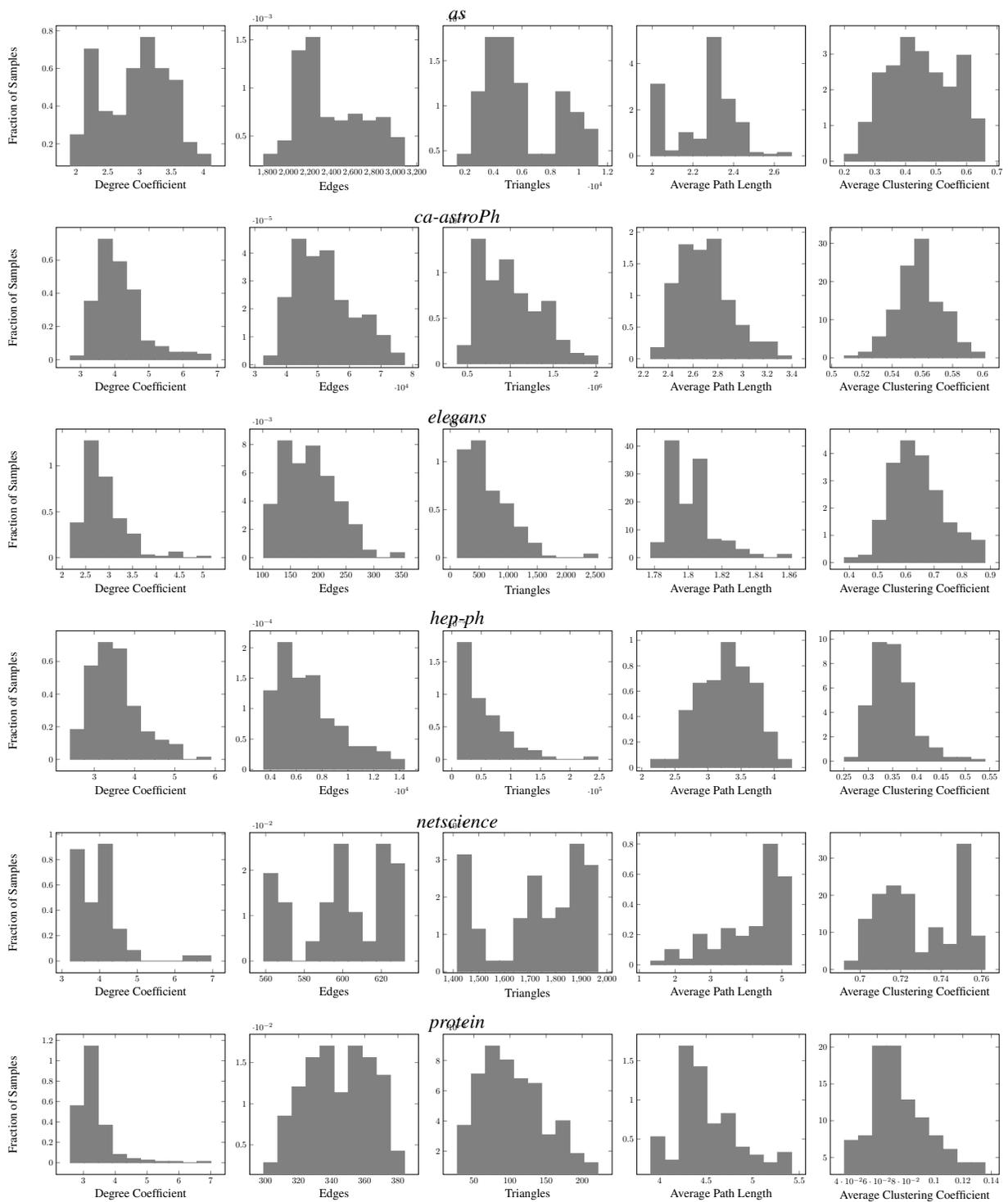
\begin{figure*}[p]
\center
%\begin{adjustbox}{width=\textwidth}
\begin{tikzpicture}[scale=0.4]
\begin{groupplot}
    [
      group style=
      {
        columns=5,
        rows = 6,
        ylabels at=edge left,
        vertical sep=2.5cm
      },
    ]
    \nextgroupplot[xlabel={\Large Degree Coefficient}, ylabel={\Large Fraction of Samples},y label style={at={(-0.2,0.5)}}]
    \addplot[hist={data=x,density},color=gray,fill]
    file {results/python_features_values/1_degree};
    \nextgroupplot[xlabel={\Large Edges},]
    \addplot[hist={data=x,density},color=gray,fill]
    file {results/python_features_values/1_edges};
    \nextgroupplot[xlabel={\Large Triangles},]
    \addplot[hist={data=x,density},color=gray,fill]
    file {results/python_features_values/1_tri};
    \nextgroupplot[xlabel={\Large Average Path Length},]
    \addplot[hist={data=x,density},color=gray,fill]
    file {results/python_features_values/1_apl};
    \nextgroupplot[xlabel={\Large Average Clustering Coefficient},]
    \addplot[hist={data=x,density},color=gray,fill]
    file {results/python_features_values/1_cc};
    \nextgroupplot[xlabel={\Large Degree Coefficient}, ylabel={\Large Fraction of Samples},y label style={at={(-0.2,0.5)}}]
    \addplot[hist={data=x,density},color=gray,fill]
    file {results/python_features_values/2_degree};
    \nextgroupplot[xlabel={\Large Edges},]
    \addplot[hist={data=x,density},color=gray,fill]
    file {results/python_features_values/2_edges};
    \nextgroupplot[xlabel={\Large Triangles},]
    \addplot[hist={data=x,density},color=gray,fill]
    file {results/python_features_values/2_tri};
    \nextgroupplot[xlabel={\Large Average Path Length},]
    \addplot[hist={data=x,density},color=gray,fill]
    file {results/python_features_values/2_apl};
    \nextgroupplot[xlabel={\Large Average Clustering Coefficient},]
    \addplot[hist={data=x,density},color=gray,fill]
    file {results/python_features_values/2_cc};
    \nextgroupplot[xlabel={\Large Degree Coefficient}, ylabel={\Large Fraction of Samples},y label style={at={(-0.2,0.5)}}]
    \addplot[hist={data=x,density},color=gray,fill]
    file {results/python_features_values/3_degree};
    \nextgroupplot[xlabel={\Large Edges},]
    \addplot[hist={data=x,density},color=gray,fill]
    file {results/python_features_values/3_edges};
    \nextgroupplot[xlabel={\Large Triangles},]
    \addplot[hist={data=x,density},color=gray,fill]
    file {results/python_features_values/3_tri};
    \nextgroupplot[xlabel={\Large Average Path Length},]
    \addplot[hist={data=x,density},color=gray,fill]
    file {results/python_features_values/3_apl};
    \nextgroupplot[xlabel={\Large Average Clustering Coefficient},]
    \addplot[hist={data=x,density},color=gray,fill]
    file {results/python_features_values/3_cc};
    \nextgroupplot[xlabel={\Large Degree Coefficient}, ylabel={\Large Fraction of Samples},y label style={at={(-0.2,0.5)}}]
    \addplot[hist={data=x,density},color=gray,fill]
    file {results/python_features_values/4_degree};
    \nextgroupplot[xlabel={\Large Edges},]
    \addplot[hist={data=x,density},color=gray,fill]
    file {results/python_features_values/4_edges};
    \nextgroupplot[xlabel={\Large Triangles},]
    \addplot[hist={data=x,density},color=gray,fill]
    file {results/python_features_values/4_tri};
    \nextgroupplot[xlabel={\Large Average Path Length},]
    \addplot[hist={data=x,density},color=gray,fill]
    file {results/python_features_values/4_apl};
    \nextgroupplot[xlabel={\Large Average Clustering Coefficient},]
    \addplot[hist={data=x,density},color=gray,fill]
    file {results/python_features_values/4_cc};
    \nextgroupplot[xlabel={\Large Degree Coefficient}, ylabel={\Large Fraction of Samples},y label style={at={(-0.2,0.5)}}]
    \addplot[hist={data=x,density},color=gray,fill]
    file {results/python_features_values/5_degree};
    \nextgroupplot[xlabel={\Large Edges},]
    \addplot[hist={data=x,density},color=gray,fill]
    file {results/python_features_values/5_edges};
    \nextgroupplot[xlabel={\Large Triangles},]
    \addplot[hist={data=x,density},color=gray,fill]
    file {results/python_features_values/5_tri};
    \nextgroupplot[xlabel={\Large Average Path Length},]
    \addplot[hist={data=x,density},color=gray,fill]
    file {results/python_features_values/5_apl};
    \nextgroupplot[xlabel={\Large Average Clustering Coefficient},]
    \addplot[hist={data=x,density},color=gray,fill]
    file {results/python_features_values/5_cc};
    \nextgroupplot[xlabel={\Large Degree Coefficient}, ylabel={\Large Fraction of Samples},y label style={at={(-0.2,0.5)}}]
    \addplot[hist={data=x,density},color=gray,fill]
    file {results/python_features_values/6_degree};
    \nextgroupplot[xlabel={\Large Edges},]
    \addplot[hist={data=x,density},color=gray,fill]
    file {results/python_features_values/6_edges};
    \nextgroupplot[xlabel={\Large Triangles},]
    \addplot[hist={data=x,density},color=gray,fill]
    file {results/python_features_values/6_tri};
    \nextgroupplot[xlabel={\Large Average Path Length},]
    \addplot[hist={data=x,density},color=gray,fill]
    file {results/python_features_values/6_apl};
    \nextgroupplot[xlabel={\Large Average Clustering Coefficient},]
    \addplot[hist={data=x,density},color=gray,fill]
    file {results/python_features_values/6_cc};
\end{groupplot}
\node (title) at ($(group c2r1.center)!0.5!(group c2r1.center)+(5cm,3.3cm)$) {{\small\em as}};
\node (title) at ($(group c2r2.center)!0.5!(group c2r2.center)+(5cm,3.3cm)$) {{\small\em ca-astroPh}};
\node (title) at ($(group c2r3.center)!0.5!(group c2r3.center)+(5cm,3.3cm)$) {{\small\em elegans}};
\node (title) at ($(group c2r4.center)!0.5!(group c2r4.center)+(5cm,3.3cm)$) {{\small\em hep-ph}};
\node (title) at ($(group c2r5.center)!0.5!(group c2r5.center)+(5cm,3.3cm)$) {{\small\em netscience}};
\node (title) at ($(group c2r6.center)!0.5!(group c2r6.center)+(5cm,3.3cm)$) {{\small\em protein}};
\end{tikzpicture}
%\end{adjustbox}
\caption{Empirical distribution of graph characteristics for the graphs listed in Table~\ref{tab:datasets}. The histogram is obtained by taking multiple random sub-graph samples and measuring five properties of the sub-graph.}
\label{fig:sampledistributions1}
\end{figure*}
The empirical distributions show that for most of the characteristics, the population of subgraphs exhibit significant variation. Moreover, the variation itself is not completely arbitrary, but in most cases, resembles standard distributions such as Gaussian or exponential.

\section{Proposed \xkpgm Model}
\label{sec:model}
%As we will show in Section~\ref{sec:results},
Existing Kronecker product based models are inadequate in capturing the natural variance that exists in graph populations. To address this issue, we propose a mixture-model based approach which allows capturing this variance. We call the proposed model \xkpgm, where the `x' signifies that all variants of \kpgm discussed in Section~\ref{sec:background} are specific instantiations of \xkpgm.

The key idea behind \xkpgm is to use $k$ ($k \ge 1$) initiator matrices of possibly different sizes. The graph generation follows similar iterative expansion as seen in other models discussed in Section~\ref{sec:background} with the difference that at any iteration one of the $k$ initiator matrices is randomly chosen for the Kronecker product by drawing a sample from a {\em Multinomial distribution} parameterized by a $k$ length probability vector ${\bm \pi}$ ($\sum_i^k\pi_i = 1$). Similar to \mkpgm, there is a parameter $l$ ($1 \le l \le k$) such that for first $l$ iterations no realization (or tying) is performed while from iteration $l+1$, parameter tying through realizations is employed.

To illustrate the impact of using multiple initiator matrices instead of a single matrix, we generate multiple synthetic graphs using different generative models. We use the following two initiator matrices:
\begin{align}
\Theta_1 & = \left(
\begin{array}{cc}
  0.95 & 0.60\\
  0.60 & 0.20
\end{array}
\right) &
\Theta_2 & = \left(
\begin{array}{cc}
  0.99 & 0.20\\
  0.20 & 0.70
\end{array}
\right)
\label{eqn:init}
\end{align}
\begin{figure}[h]
\centering
\begin{tikzpicture}[scale=0.8]
  \begin{groupplot}
    [
      group style=
      {
        columns=1,
        rows = 2,
        ylabels at=edge left,
        vertical sep=2.5cm
      },
    ]
    \nextgroupplot[xlabel={\Large Tying Parameter $l$}, ylabel={\Large Average Number of Edges}]
    \addplot+[/pgfplots/error bars/.cd,x dir=none,y dir=both,y explicit] table [x = a, y = b,y error=e]{data.dat};
    \addlegendentry{\mkpgm ($\Theta_1$)}
    \addplot+[/pgfplots/error bars/.cd,x dir=none,y dir=both,y explicit] table [x = a, y = j,y error=m]{data.dat};
    \addlegendentry{\mkpgm ($\Theta_2$)}
    \nextgroupplot[xlabel={\Large Tying Parameter $l$},ylabel={\Large Average Number of Edges},ymin = 0, ymax = 8000]
    \addplot+[/pgfplots/error bars/.cd,x dir=none,y dir=both,y explicit] table [x = a, y = c,y error=f]{data.dat};
    \addlegendentry{\xkpgm ($\pi = 0.5$)}
    \addplot+[/pgfplots/error bars/.cd,x dir=none,y dir=both,y explicit] table [x = a, y = i,y error=l]{data.dat};
    \addlegendentry{\xkpgm ($\pi = 0.8$)}
\end{groupplot}
\end{tikzpicture}
\caption{Variation of {\em number of edges} for synthetic graphs with $2^{10}$ nodes generated by \mkpgm and \xkpgm for varying levels of tying ($l$). Note that $l=1$ indicates purely tied model (\tkpgm) and $l=10$ indicates purely untied model (\kpgm). The error bars are computed by generating 2000 samples for each value of $l$.}
\label{fig:simulations}
\end{figure}
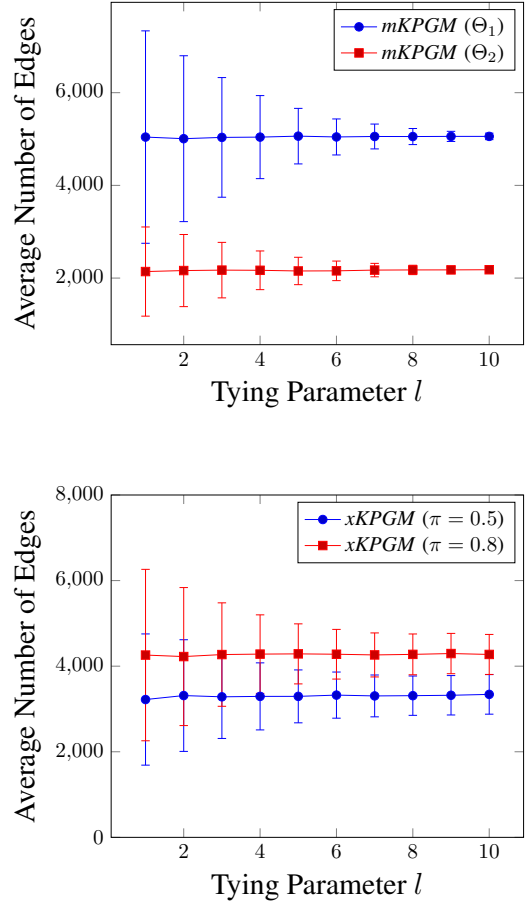
These exact matrices have been used in previous papers~\cite{Moreno:2010} for conducting empirical simulations for \mkpgm and other variants. The mean and the variance for the number of edges across the simulations is shown in Figure~\ref{fig:simulations} which illustrates that the variability (in terms of the number of edges) for samples generated by \mkpgm is low for higher values of $l$ ($\ge 6$). In fact, the variance becomes negligible for \kpgm ($l = 10$). For very low values of $l$ the variability is very high. On the other hand, \xkpgm generates samples with much more stable variability across $l$. We observe that for \xkpgm, the variance is generally higher when $\pi = 0.5$, though it also depends on the individual initiator matrices. We observe similar behavior for other properties (clustering coefficient, etc.) as well.
\subsection{\xkpgm Sampling Algorithm}
The steps for sampling a graph in \xkpgm are shown in Algorithm~\ref{alg:sampling}. Note that the algorithm needs following inputs: $k$ initialization matrices, ${\Theta_1,\Theta_2,\ldots,\Theta_k}$, a $k$ length vector ${\bm \pi}$, the number of iterations $n$, and the level of tying $l$. The number of rows (and columns) for each matrix $\Theta_i$ is denoted as $b_i$.
\subsection{Interchangeability in Kronecker Products}
\label{subsec:interchange}
We make use of the following important result~\cite{Henderson:1981} for Kronecker matrices. For two matrices $A$ and $B$:
\begin{equation}
A \otimes B = M(B \otimes A)N
\label{eqn:7}
\end{equation}
where $M$ and $N$ are {\em permutation matrices} which merely change the order of rows and columns, respectively. In other words, the above result means that the Kronecker product is {\em commutative} as long as we allow changing of labels of nodes, which is irrelevant in our case. Since in Algorithm~\ref{alg:sampling}, the order of rows and columns is not important, one can use this result in the purely untied setting.
\begin{obs}
In Algorithm~\ref{alg:sampling}, the graph generated using any arbitrary sequence of initialization matrices is equivalent to the following canonical sequence:
\[
\underbrace{(\Theta_1 \otimes \Theta_1 \ldots)}_{n_1\text{ times}}\otimes\underbrace{(\Theta_2 \otimes \Theta_2 \ldots)}_{n_2\text{ times}}\otimes \ldots \otimes \underbrace{(\Theta_k \otimes \Theta_k \ldots)}_{n_k\text{ times}}
\]
\label{obs:2}
\end{obs}
The above observation allows us to arrive at several important results later, because it simplifies the analysis, especially, in the untied setting for \xkpgm, where arbitrary sequences of initialization matrices can be represented using the above canonical sequence. In fact, the next result follows from this observation.
\subsection{Generating Graphs of Arbitrary Sizes}
The input $n$ determines the eventual size of the graph in terms of the number of nodes ($\vert {\bf V}\vert$). For a simple case where all initialization matrices are of the same size ($b_i = b, \forall i$), the size of the graph will be $b^{n}$. In general, expected number of nodes will be $\prod_{i=1}^kb_i^{n\pi_i}$.

Note that existing Kronecker based models are limited to generating graphs with number of nodes as exact powers of $b$ (typically set to 2 or 3). On the other hand, \xkpgm can generate graphs with number of nodes, which in principle, need not be restricted to the exact powers of $b$. Thus, by using initialization matrices of varying sizes, \xkpgm can generate graphs with more number of possibilities for the number of nodes. For example, if we use two initialization matrices of sizes 2 and 3 and if $N$ can be factorized as $2^a3^b$, then one can set $\pi_1 = \frac{a}{a+b}$ and $\pi_2 = \frac{b}{a+b}$ and generate a graph using Algorithm~\ref{alg:sampling} with $N$ expected number of nodes.
%\begin{algorithm}
%\caption{Graph Generation Algorithm for \xkpgm}\label{alg:sampling}
%\begin{algorithmic}[1]
%\Procedure{Euclid}{$a,b$}\Comment{The g.c.d. of a and b}
%  \State $r\gets a\bmod b$
%  \While{$r\not=0$}\Comment{We have the answer if r is 0}
%  \State $a\gets b$
%  \State $b\gets r$
%  \State $r\gets a\bmod b$
%  \EndWhile\label{euclidendwhile}
%  \State \textbf{return} $A$
%\EndProcedure
%\end{algorithmic}
%\end{algorithm}
\begin{algorithm}[h]
{\small
%\SetLine
%\DontPrintSemicolon
\KwIn{$\Theta_1,\Theta_2,\ldots,\Theta_k,{\bm \pi},n,l$}
\KwOut{Adjacency matrix $A$}
$\mathcal{P} \leftarrow 1$\\
\tcp{Untied Phase}
\ForEach{$t = 1$ to $l$}{
  $i \sim Multinomial({\bm \pi})$\\
  $\mathcal{P} \leftarrow \mathcal{P} \otimes \Theta_i$\\
}
\tcp{Tied Phase}
\ForEach{$t=l+1$ to $n$}{
  $A \leftarrow R(\mathcal{P})$\tcp*{$R$ - Realization}
  $i \sim Multinomial({\bm \pi})$\\
  $\mathcal{P} \leftarrow A \otimes \Theta_i$\\
}
$A \leftarrow R(\mathcal{P})$\\
\Return $A$
\caption{Graph Generation Algorithm for \xkpgm}\label{alg:sampling}
}
\end{algorithm}
%and a $k$ length vector ${\bm \pi}$ such that $\sum_{i}\pi_i = 1$ and $\pi_i$ denotes the probability of selecting $i^{th}$ initiator 
%\subsection{Parameters of \xkpgm}
%\xkpgm is parameterized by 
\subsection{Relationship to Existing Models}
If $k=1$ and $l=1$, \xkpgm is equivalent to \tkpgm. Similarly, for $k=1$ and $l=n$, \xkpgm is equivalent to \kpgm. For $k=1$ and for all other values of $l$ ($1 \le l < n$) \xkpgm is equivalent to \mkpgm. Thus all existing Kronecker product based graph models are specific instantiations of \xkpgm.
\begin{table*}[t]
\centering
{\small
\begin{tabular}{|c|c|p{3in}|}
\hline
Model & $\mathbb{E}[{\bf E}]$ & $var[{\bf E}]$\\
\hline\hline
\kpgm & $S_{\Theta}^n$& $S_{\Theta}^n - S_{\Theta^2}^n$\\
\hline
\tkpgm & $S_{\Theta}^n$& $S_\Theta^{n-1}(S_\Theta^n - 1)\frac{S_\Theta - S_{\Theta^2}}{S_\Theta - 1}$\\
\hline
\mkpgm & $S_{\Theta}^n$& $S_\Theta^{n-1}(S_\Theta^{n-l} - 1)\frac{S_\Theta - S_{\Theta^2}}{S_\Theta - 1} + (S_\Theta^l - S_{\Theta^2}^l)S_\Theta^{2(n-l)}$\\
\hline
\xkpgm (purely tied) & $(\pi_1S_{\Theta_1}+\pi_2S_{\Theta_2})^n$& $(\pi_1(S_{\Theta_1} - S_{\Theta_1^2}) + \pi_2(S_{\Theta_2} - S_{\Theta_2^2}) + \pi_1\pi_2(S_{\Theta_1} - S_{\Theta_2})^2)\sum_{j=1}^{2(n-1)}(\pi_1S_{\Theta_1}+\pi_2S_{\Theta_2})^j$\\
\hline
\xkpgm (untied) & $S_{\Theta_1}^{\pi_1n}S_{\Theta_2}^{\pi_2n}$& $S_{\Theta_1}^{\pi_1n}S_{\Theta_2}^{\pi_2n} - S_{\Theta_1^2}^{\pi_1n}S_{\Theta_2^2}^{\pi_2n}$\\
\hline
\end{tabular}
\caption{Comparison of $\mathbb{E}[{\bf e}]$ and $var[{\bf e}]$ for Kronecker models. $S_\Theta$, $S_{\Theta_1}$ and $S_{\Theta_2}$ denote the sum of values of $\Theta$, $\Theta_1$ and $\Theta_2$. $S_{\Theta^2}$, $S_{\Theta^2_1}$ and $S_{\Theta^2_2}$ denote the sum of squares of values of $\Theta$, $\Theta_1$ and $\Theta_2$.}%Expressions for \xkpgm are provided for $k=2$ though the same expressions can be generalized for more than 2 initiator matices.}
\label{tab:comparison}
}
\end{table*}

Kronecker product models have been analyzed in terms of the expected number of edges for the generated samples. In Section~\ref{sec:properties}, we provide similar expressions for other graph properties as well. In Table~\ref{tab:comparison}, we list the expressions for the expectation and variance for the number of edges for \xkpgm and compare them with other models. The derivation is omitted in the interest of space though the key is to use the Observation~\ref{obs:2} along with the fact that the expectation of an edge between a pair of nodes is equal to the corresponding value in the final stochastic matrix.

\section{Properties of \xkpgm Graphs}
\label{sec:properties}
In this section we provide expressions for expectations of four key properties for graphs, viz.,
\begin{inparaenum}[i)]
\item edges,
\item hairpins (or 2-stars),
\item tripins (or 3-stars), and
\item triangles.
\end{inparaenum}
These four properties provide an understanding of local structure of the samples. Moreover, we will be using the expressions to estimate the model parameters in Section~\ref{sec:estimation}. For simplicity we assume that all $k$ initiator matrices are of size $2 \times 2$ and are of the following form:
%\begin{align}
$\Theta_k = \left(
\begin{array}{cc}
  a_k & b_k\\
  b_k & c_k
\end{array}
\right) $.
%\label{eqn:init1}
%\end{align}
The derivation for expected values in the purely untied ($l = n$) setting closely follows the line of argument taken for pure \kpgm~\cite{Gleich:2012} (Section 3) and makes use of Observation~\ref{obs:2}. The expressions are provided in Table~\ref{tab:expressions}.
\begin{table*}
\centering
{\footnotesize
\begin{tabular}{ccl}
\toprule
$2\mathbb{E}[{\bf E}]$ & $=$ & $\prod_{i=1}^n(a_i+2b_i+c_i)^{\pi_in} - \prod_{i=1}^n(a_i + c_i)^{\pi_in}$\\
\midrule
$2\mathbb{E}[{\bf H}]$ & $=$ & $\prod_{i=1}^n((a_i + b_i)^2 + (b_i + c_i)^2)^{\pi_in} - 2\prod_{i=1}^n(a_i(a_i + b_i)+c_i(c_i + b_i))^{\pi_in} - \prod_{i=1}^n(a_i^2 + 2b_i^2 + c_i^2)^{\pi_in} + 2\prod_{i=1}^n(a_i^2 + c_i^2)^{\pi_in}$\\
\midrule
$6\mathbb{E}[{\bf T}] $ & $=$ & $\prod_{i=1}^n((a_i+b_i)^3+(b_i+c_i)^3)^{\pi_in} - 3\prod_{i=1}^n(a_i(a_i+b_i)^2+c_i(b_i+c_i)^2)^{\pi_in}$\\
& $-$ &$3\prod_{i=1}^n(a_i^3+c_i^3+b_i(a_i^2+c_i^2)+b_i^2(a_i+c_i)+2b_i^3)^{\pi_in} + 2\prod_{i=1}^n(a_i^3+2b_i^3+c_i^3)^{\pi_in}$\\
& $+$ &$5\prod_{i=1}^n(a_i^3+c_i^3+b_i^2(a_i+c_i))^{\pi_in}+4\prod_{i=1}^n(a_i^3+c_i^3+b_i(a_i^2+c_i^2))^{\pi_in} - 6\prod_{i=1}^n(a_i^3+c_i^3)^{\pi_in}$\\
\midrule
$6\mathbb{E}[{\bm \Delta}]$ & $=$ & $\prod_{i=1}^n(a_i^3 + 3b_i^2(a_i+c_i) + c_i^3)^{\pi_in}- 3\prod_{i=1}^n(a_i(a_i^2+b_i^2)+c_i(b_i^2 + c_i^2))^{\pi_in} + 2\prod_{i=1}^n(a_i^3 + b_i^3)^{\pi_in}$\\
\bottomrule
\end{tabular}
}
\caption{Expressions for ${\bf E}$ - num. edges, ${\bf H}$ - num. hairpins, ${\bf T}$ - num. tripins and ${\bm \Delta}$ - num. of triangles}
\label{tab:expressions}
\end{table*}

Note that the expression for $\mathbb{E}[{\bf E}]$ here is slightly different from a similar expression in Table~\ref{tab:comparison}. The reason is that in Table~\ref{tab:expressions} we do not count the self-loops and count each edge only once.

%\section{Theoretical Analysis of \xkpgm}
%\input{analysis}
\section{Parameter Estimation}
\label{sec:estimation}
The parameters of \xkpgm are the elements in the $k$ initiator matrices and the vector ${\bm \pi} = \pi_1,\pi_2,\ldots, \pi_k$. We describe a {\em method of moments} based approach to estimate these parameters from a given observed graph. The graph is characterized using a set of statistics (or {\em moments}). In this paper we have used the same moments as used in the past for \kpgm estimation (edges, hairpins, tripins, and triangles)~\cite{Gleich:2012}. The idea is to find the model parameters such that the expected moments for the model match closely with the moments computed from the observed graph within an optimization procedure. Each moment is denoted as $F_i$ and the computed moment for an observed graph $\mathcal{G}$ is denoted as $F_i^*$. The estimation method searches for the parameters $\Theta_1,\Theta_2,\ldots, {\bm \pi}$ which minimize the following objective function:
\begin{equation}
f({\Theta},{\bf F}^*) = \sum_{i=1}^{|{\bf F}|}w_i\left(\frac{F_i^* - \mathbb{E}[F_i|\Theta]}{F_i^*}\right)^2
\label{eqn:3}
\end{equation}
where $w_i$ indicates a weight assigned to the $i^{th}$ moment. In this paper we have used equal weights, however, they can be used to assign more emphasis on certain moments. Note that the above objective function can be easily extended to learn from multiple graphs by replacing the observed moment $F_i$ in~\eqref{eqn:3} with the average moment value across all $p$ observed graphs, i.e., $\widetilde{F}_i = \frac{1}{p}\sum_{\mathcal{G}}F_{i\mathcal{G}}$.

For \xkpgm in a purely untied setting ($l = n$), we have expressions for each of the moments as functions of the parameters (See Table~\ref{tab:expressions}). We plugin these expressions into a gradient based optimizer (e.g., {\em fmincon} function available in Matlab) to get the optimal parameter values.

Note that the above estimation algorithm can be extended to more general settings of purely untied \xkpgm such as using initiator matrices of larger sizes ($> 2$ and using initiator matrices with different sizes, by deriving the appropriate expressions similar to the ones listed in Table~\ref{tab:expressions}. For the tied setting ($l < n$), obtaining such closed form expressions is challenging. However, one could use the {\em simulated method of moments} approach which was originally proposed for \mkpgm~\cite{Moreno:2013}.

\section{Computational Complexity}
\label{sec:complexity}
Generating a graph of size $N$ using the learnt initiator matrices using Algorithm~\ref{alg:sampling} requires $n = \log_2{N}$ Kronecker products and $l$ intermediate realization followed by $N^2$ bernoulli trials for the final realization. In the original \kpgm paper~\cite{Leskovec:2010}, the authors propose a sampling strategy which is linear in the number of edges, however, that method requires one to specify the number of edges desired in the graph. For \xkpgm the generation is dominated by the final $O(N^2)$ realization step. However, given that the realizations are independent, efficient distributed implementations can be devised.

The parameter estimation using the method of moments allows scalable training. The only compute intensive phase is the computation of the moments for the observed graph, which is done once. The objective function can be computed in $O(1)$ time, independent of the size of the observe graph. In contrast, the simulated method of moments, used by \mkpgm, requires computation of the moments for the generated graph multiple times at each iteration of the gradient descent, making it unscalable to large graphs. The number of iterations required by the {\em fmincon} minimizer will depend on the number of model parameters, which increases linearly with the number of initiator matrices ($k$).

\section{Experimental Results}
\label{sec:results}
We present results on several publicly available network data sets (See Table~\ref{tab:datasets}). The evaluation has two objectives. We first show that the parameters learnt for \xkpgm using the algorithm sketched in Section~\ref{sec:estimation} allow a more accurate modeling of the observed graph, compared to other Kronecker based models.
\subsection{Parameter Estimation}
For each graph in Table~\ref{tab:datasets}, we estimate parameters using Matlab {\em fmincon} with 5000 initial values with constraints that elements of the initiator matrices are between 0 and 1 and $a_i > c_i$. We train \xkpgm with two initiator matrices of size $2\times 2$ and other models (\kpgm, \tkpgm, \mkpgm) using one initator matrix of size $2\times 2$, using the objective function in \eqref{eqn:3} and the four moments listed in Table~\ref{tab:expressions}. For \mkpgm and \tkpgm, we use implementation provided by the authors~\cite{Moreno:2013}\footnote{\url{https://www.cs.purdue.edu/homes/smorenoa/mKPGM.zip}}. For \kpgm, we used the direct optimization method~\cite{Gleich:2012}. 
%Note that the authors had provided three estimation procedures, but the objective function values were the best for the direct method.
\subsection{Matching Moments}
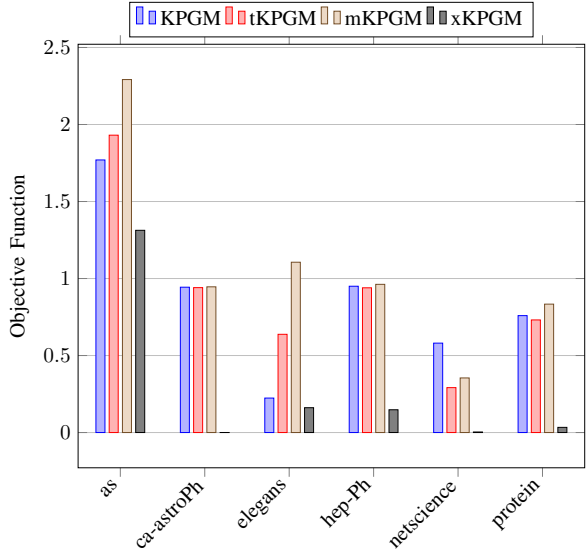
\begin{figure}[h]
\centering
\begin{tikzpicture}[scale=0.8]
\begin{axis}[
%    symbolic x coords={as,astroPh,elegans,hepPh,netscience,protein,sd},
    width=10cm,
    xtick={1,2,3,4,5,6},xticklabels={as,ca-astroPh,elegans,hep-Ph,netscience,protein},
    x tick label style={rotate=45, anchor=east},
    ylabel=Objective Function,
    ymajorgrids = true,
    xmajorgrids = false,
    legend style={at={(0.5,1.1)},
      anchor=north,legend columns=4,},
    ybar,% interval=1,
    bar width=0.15cm,
]
\addplot 
coordinates{
(1,1.7698)
(2,0.9436)
(3,0.2238)
(4,0.9499)
(5,0.5805)
(6,0.7591)
};
\addplot
coordinates{
(1,1.9307)
(2,0.9409)
(3,0.6378)
(4,0.9397)
(5,0.2916)
(6,0.7311)
};
\addplot
coordinates{
(1,2.2918)
(2,0.9461)
(3,1.1058)
(4,0.9622)
(5,0.3545)
(6,0.8337)
};
\addplot
coordinates{
(1,1.313)
(2,0.0003)
(3,0.1613)
(4,0.1479)
(5,0.0035)
(6,0.0336)
};
\legend{KPGM,tKPGM,mKPGM,xKPGM}
\end{axis}
\end{tikzpicture}
\caption{Comparing \xkpgm with other models in terms of the objective function value obtained after training.}
\label{fig:momentcomparison}
\end{figure}

The objective function values (see~\eqref{eqn:3}) for the different methods are shown in Figure~\ref{fig:momentcomparison}. 
%The estimated parameters and the objective function values (see~\eqref{eqn:3}) are reported in Table~\ref{tab:results}. The last column indicates the value of the objective function for each algorithm. 
For every data set, \xkpgm gives the best estimation in terms of the objective function, especially for {\em ca-astroPh}, {\em netscience}, {\em protein}, and {\em hep-ph} data sets. The relative performance of other models varies across the different data sets. In general, we observe that the objective function value improves as the tying level ($l$) decreases and is best when $l = 1$, i.e., \tkpgm.

\subsection{Impact of Number of Initiator Matrices}
\begin{figure}[h]
\centering
\begin{tikzpicture}[scale=0.9]
\begin{axis}[
    xtick={0,1,...,4},
    ylabel=Objective Function,
    xlabel=Number of Initiator Matrices - $k$,
    legend style={at={(0.5,1.1)},
      anchor=north,legend columns=3,},
]
\addplot[ thick,mark=otimes]
coordinates{
(1,1.7698)
(2,1.3130)
(3,1.1251)
(4,1.0674)
};
\addplot[ thick, mark=square]
coordinates{
(1,0.5805)
(2,0.0035)
(3,0.0090841)
(4,0.0076623)
};
\addplot[ thick,mark=triangle]
coordinates{
(1,0.7591)
(2,0.0336)
(3,0.0035467)
(4,0.0035467)
};
\legend{as,netscience,protein}
\end{axis}
\end{tikzpicture}
\caption{Performance of \xkpgm using different number of initiator matrices ($k$). Note that $k=1$ is \kpgm.}
\label{fig:kimpact}
\end{figure}
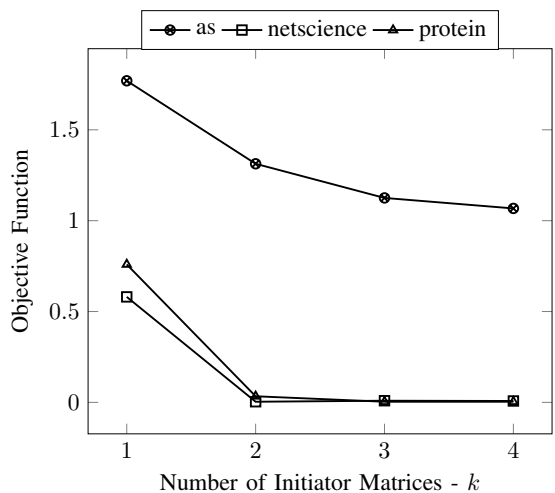

To understand how the choice of $k$, i.e., the number of initiator matrices, impacts the performance of the model, we compared the objective function obtained by training the \xkpgm model using different values of $k$. The results for three graphs are shown in Figure~\ref{fig:kimpact}. In all three cases, increasing $k$ from 1 to 2 results in a significant improvement in the objective function. For two of the graphs (netscience and protein), increasing $k$ beyond 2 does not show any further improvements. However, for the $as$ graph, higher values of $k$ show improvement in the objective function.

%In terms of individual moments, the number of triangles was the hardest to match and was consistently under-represented in the estimated models, including \xkpgm.
\subsection{Variability of Generated Samples}
\begin{figure}[t]
\centering
\begin{subfigure}{0.52\textwidth}
   \includegraphics[width=\linewidth]{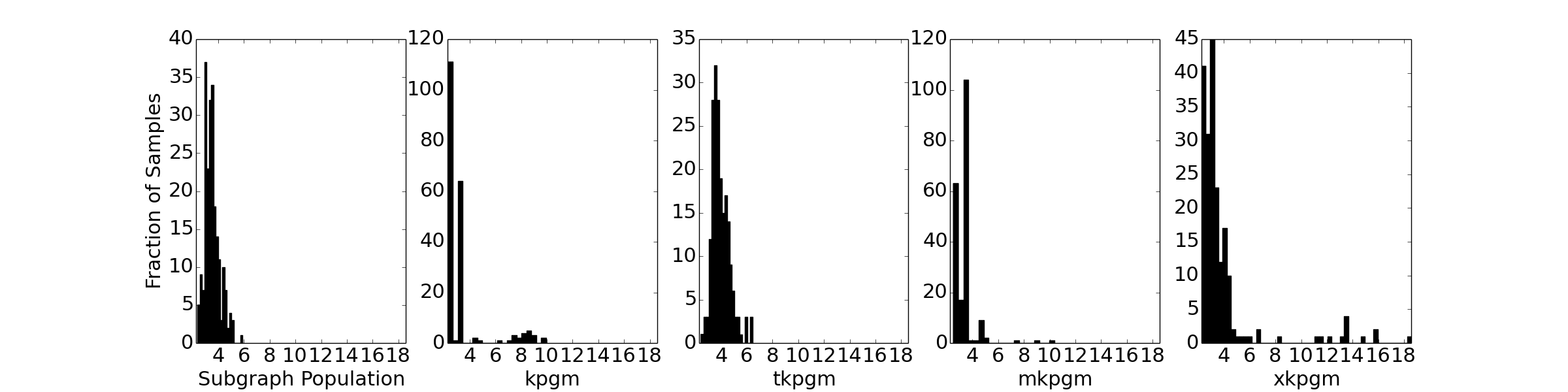}
  \caption{Power Law Coefficient of Degree Distribution}
\end{subfigure}
%\begin{subfigure}{0.52\textwidth}
%   \includegraphics[width=\linewidth]{comparison_edges.png}
%  \caption{Number of Edges}
%\end{subfigure}
%\begin{subfigure}{0.52\textwidth}
%   \includegraphics[width=\linewidth]{comparison_triangles.png}
%  \caption{Number of Triangles}
%\end{subfigure}
\begin{subfigure}{0.52\textwidth}
   \includegraphics[width=\linewidth]{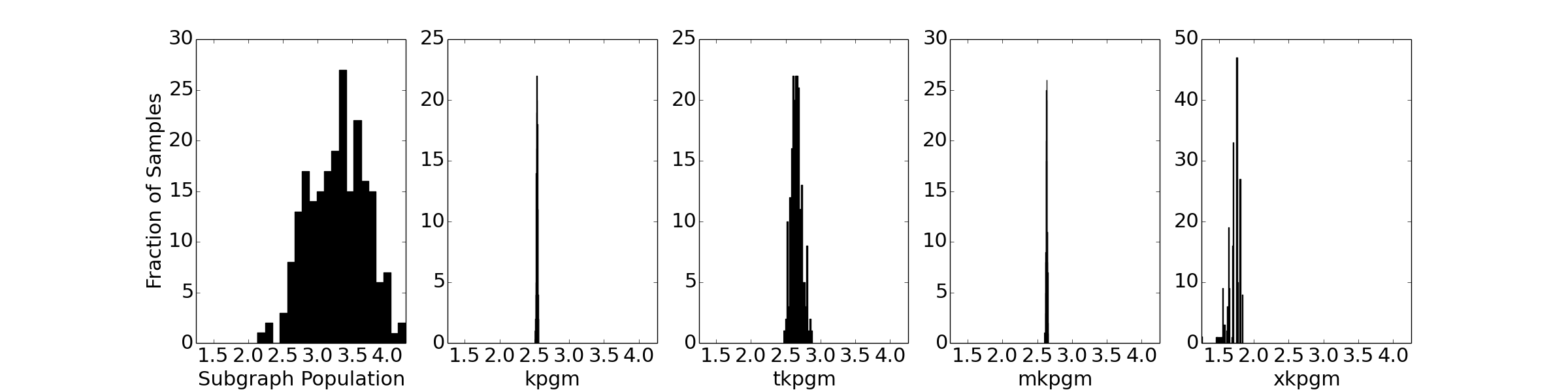}
  \caption{Average Path Length}
\end{subfigure}
\begin{subfigure}{0.52\textwidth}
   \includegraphics[width=\linewidth]{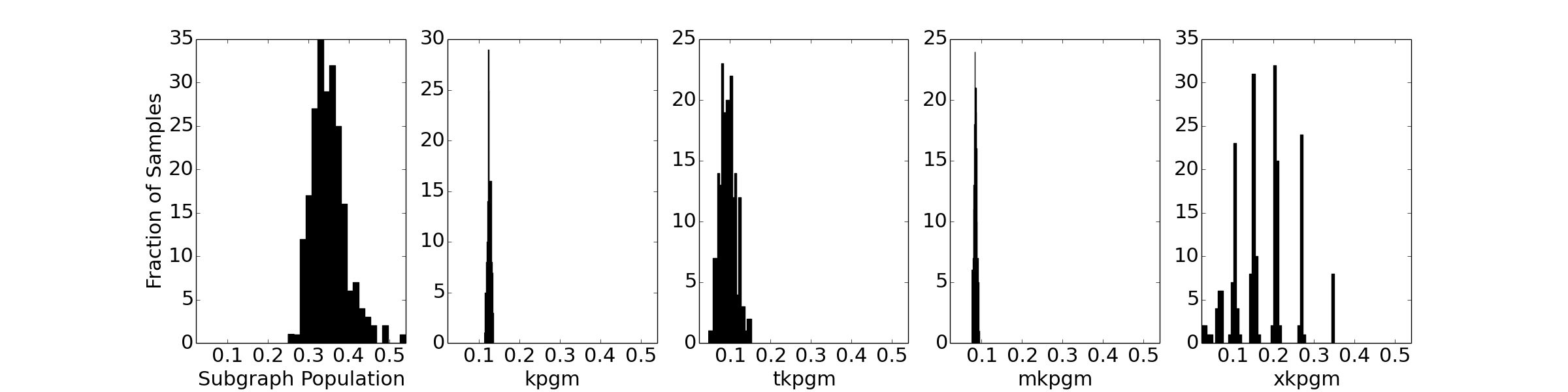}
  \caption{Average Clustering Coefficient}
\end{subfigure}
\caption{Distribution of network characteristics for samples using the models trained on {\em cit-hepPh} data. The first plot in each row is the distribution for the subgraph population discussed in Section~\ref{sec:variability}.}
\label{fig:variabilityresults}
\end{figure}

The second set of experiments attempt to understand the variability in the samples generated by the \xkpgm and other Kronecker models. For each model, we use the estimated parameters and generate 200 samples with the same sample size that was used to generate the subgraphs in Section~\ref{sec:variability}. For each sample we measure five salient properties similar to Figure~\ref{fig:sampledistributions1} and plot the distribution across the samples. The distributions for {\em cit-hepPh} is shown in Figure~\ref{fig:variabilityresults}\footnote{Results for other data sets available at: \protect\url{http://www.cse.buffalo.edu/~chandola/research/bigdata2015graphs/allresults.pdf} due to lack of space.}.
%As noted earlier, \kpgm and \mkpgm do not capture the variability that is observed in the subgraph population, whereas \tkpgm shows higher variability, even though \mkpgm was proposed to achieve bounded variance. However, 
In the results shown in Figure~\ref{fig:variabilityresults}, both \kpgm and \mkpgm exhibit almost negligible variability for almost all of the network properties. In fact, it is evident that the variability increases as we decrease $l$ from $n$ (\kpgm) to $1$ (\tkpgm). \xkpgm shows most variability, and in many cases approximately matches the variability of the subgraph population (degree distribution, average clustering coefficient). 
%The variability induced by \xkpgm exhibits a two mode distribution which aligns with the fact that we chose two initiator matrices for the model. We expect that for $k > 2$, we will observe $k$ modes in the distribution.
\subsection{Comparison with \bter}
As noted earlier, the \bter model allows generating a synthetic graph which approximates an observed graph. However, by design, \bter can generate graphs of approximately the same size (nodes and edges) as the observed graph. Using the available implementation\footnote{\url{http://www.sandia.gov/~tgkolda/feastpack}}, we observed that while \bter matches the degree distribution and the clustering coefficient, the generated graphs show minimal variation for a set of samples generated using the same observed graph. For example, for the {\em ca-astroPH} graph, 200 simulations using \bter resulted in graphs with number of edges in the range: $196156 \pm 208$ and the average clustering coefficient in the range: $0.3265 \pm 0.0013$.

\section{Conclusions}
We propose a new generative model, \xkpgm, for graphs based on mixture of Kronecker models. The model induces variability by incorporating multiple initiator matrices strung together using a random multinomial trial, while retaining the strong features of \kpgm, such as scaling to massive graphs. We provide analytical expressions for various characteristics of the generated graphs which allows us to estimate the parameters using a method of moments approach. The mixture approach allows us to devise a scalable method of moments based learning method (similar to \kpgm) while achieving better variance (similar to \mkpgm).

We show, both analytically and through experiments, that \xkpgm is able to learn a model which best matches the moments of the observed graph (See Figure~\ref{fig:momentcomparison}) and also induces variability (See Figure~\ref{fig:variabilityresults}) which aligns with the natural variability observed in real graphs (See Figure~\ref{fig:sampledistributions1}). To better evaluate the variability of generative graph models we have come up with a sub-graph based method. Using this method we show that the proposed model provides a robust variance across multiple graph properties. We also provide comparisons with the state of art generative models and show that \xkpgm outperforms them both in terms of matching the graph properties and the variance in the population.

%A key strength of \xkpgm is its flexibility. On one hand, all existing Kronecker models are specific instantiations of \xkpgm and on the other hand, one can derive more sophisticated models by using more initiator matrices, using matrices of varying sizes, and by using different levels of tying. In fact, as shown in Figure~\ref{fig:kimpact}, using higher number of initiator matrices resulted in a better match with the graph properties for several graphs.
%Another possible extension would be to consider other moments such as {\em four circles}, etc., since the moments considered in this paper, i.e., edges, pins, stars represent a single class of features. 

%An obvious limitation of \xkpgm, and other Kronecker product based models is that the number of nodes in the generated graph are a power of 2. Since \xkpgm uses multiple matrices of varying sizes, one can span a wider range of graph sizes. In future, we plan to incorporate the ability to generate graphs with arbitrary number of nodes in \xkpgm. Another limitation is that the variance estimates are made using a population of subgraphs. In future we plan to model actual populations of graphs using \xkpgm. 

\section*{Acknowledgements}
This work was supported by NSF grant CNS:1409551. Access to computing resources was facilitated by an AWS in Education Grant award.

%\bibliographystyle{IEEEtran}
%\bibliography{refs.bib}

\end{document}